\documentclass{aastex631}
\usepackage{amsmath}

\newcommand\pennstate{Department of Astronomy \& Astrophysics and \\
Center for Exoplanets and Habitable Worlds and \\
Penn State Extraterrestrial Intelligence Center \\
525 Davey Laboratory \\
The Pennsylvania State University \\
University Park, PA, 16802, USA}

\newcommand\nasagoddard{
NASA Goddard Space Flight Center \\
Greenbelt, Maryland, USA
}

\begin{document}

\title{The Abundance of Belatedly Habitable Planets and Ambiguities in Definitions of the Continuously Habitable Zone}

\author[0000-0003-3989-5545]{Noah W.\ Tuchow}
\affil{\nasagoddard}
\affil{\pennstate}

\author[0000-0001-6160-5888]{Jason T.\ Wright}
\affil{\pennstate}

\begin{abstract}
    A planet's history dictates its current potential to host habitable conditions and life. The concept of the Continuously Habitable Zone (CHZ) has been used to define the region around a star most likely to host planets with long-term habitability. However, definitions of the CHZ vary in the literature and often conflict with each other. Calculating the fraction of habitable zone planets in the CHZ as a function of stellar properties, we find that the quality of a star as a host for planets with long-term habitability and biosignatures depends strongly on the formulation of the CHZ used. For instance, older M stars are either excellent or sub-optimal hosts for CHZ planets, depending on whether one's definition of habitability prioritizes the total time spent in the habitable zone or the continuity of habitable conditions from the delivery of volatiles to its current age.
    
    In this study, we focus on Belatedly Habitable Zone (BHZ) planets, i.e.\ planets which enter the habitable zone after formation due to the evolution of their host star. We find that between $\sim$29-74\% of planets in the habitable zone belong to this class of BHZ planets depending on the timescale for the delivery of volatiles. Whether these planets can retain their volatiles and support habitable conditions is unclear. Since BHZ planets comprise a large portion of the planets we expect to survey for biosignatures with future missions, the open question of their habitability is an important factor for mission design, survey strategies, and the interpretation of results.
    
\end{abstract}

\section{Introduction}

Continuous and sustained habitability is an important factor when considering which planets may be the strongest candidates in the search for life. One would like to determine whether the planets that we discover have hosted temperate surface conditions long enough to allow for the emergence of life and biosignatures. The search for life typically focuses on  planets in the habitable zone, which provides a general location where planets with Earth-like climate feedback cycles could support liquid water on their surfaces \citep{Kasting1993hz,Kopparapu2013}. However, this does not imply  that all planets located in the habitable zone are actually able to support temperate surface conditions, nor does it imply that these planets necessary possess a large inventory of water.  Therefore, one should not consider all planets in the habitable zone to be equally likely to support  habitable conditions and biosignatures. Because habitability is likely a path-dependent process, determined by both the evolution of a planet's geophysical properties as well as the evolution of its host star \citep{Lenardic2016,Lenardic2021}, knowledge that a planet is currently located in the habitable zone only provides limited information as to whether it has been habitable for a long enough duration to be a candidate to host biosignatures.

\subsection{Continuous Habitability}
Stars evolve over the course of their lifetimes in terms of their luminosity and effective temperatures, and this means that their habitable zones also change in time.
Past studies have recognized the importance of the considering the duration that planets spend in the habitable zone, leading to the concept of the Continuously Habitable Zone (CHZ). There are multiple formulations for the CHZ in the literature. One definition of the CHZ is the region of the habitable zone occupied by planets that have remained in the habitable zone from the onset of habitability to the current age of the star \citep{Tuchow2020}. In this paper we shall refer to this formulation as the \textit{sustained CHZ}.
In the literature, one often sees an alternative definition of the CHZ where one considers planets which spend more than a fixed amount of time in the habitable zone to be continuously habitable, without consideration of whether planets originated in the habitable zone. We shall refer to this alternative definition of the continuously habitable zone as the \textit{fixed duration CHZ}. As in the case of the \citeauthor{Tuchow2020} sustained CHZ, where the time for the onset of habitability is a free parameter, for the fixed duration CHZ, the duration beyond which planets are considered to be continuously habitable is a free parameter. This duration ranges in the literature between values of 2 Gyr \citep{Truitt2020} to 4 Gyr \citep{Hart1979}. 

Alongside the concept of continuous habitability, there is also the opposite concept of belated habitability. Belatedly habitable planets originate outside the habitable zone and enter it later in time, either due to the evolution of their host star or due to planetary migration \citep{Tuchow2021}. This category of planets occupies the regions of the habitable zone outside of the sustained CHZ, referred to as the inner and outer Belatedly Habitable Zones (BHZ). 
In this study, we shall primarily focus on BHZ planets that enter the habitable zone due to the the easily modelled effects of stellar evolution (as opposed to orbital migration which is much more stochastic and less readily predictable). Planets in these regions of the habitable zone encounter unique obstacles to habitability, and it is unclear whether they could support habitable conditions. 

Broadly speaking, planets in the BHZ fall into two general classes based on their evolutionary histories. 
Planets originating interior to the the habitable zone are subject to intense stellar fluxes earlier in their histories, and may be desiccated by the time they enter the habitable zone \citep{Barnes2013}. 
This class of inner-BHZ planets should be common around M dwarf stars, which dim on the premain sequence for tens to hundreds of millions of years, such that most of the planets currently in the habitable zones of main sequence M stars will have originated interior to the habitable zone  \citep{Luger2015,Tuchow2021}. The habitability of such planets would depend on their initial inventory of volatiles and whether they can preserve them during an extended period of intense instellation \citep{lustig_yaeger2019}. If these planets could maintain a substantial reservoir of volatiles in their mantles, then through outgassing they may be able to support surface liquid water after entering the habitable zone \citep{moore2020,Barth2021}.  
On the opposite edge of the habitable zone, planets that start too far away from their stars will likely enter the habitable zone in a globally glaciated state with a high ice albedo, making them difficult to thaw even at much higher instellations \citep{Kasting1993hz,Kadoya2019}. The ability of this class of outer-BHZ planets to thaw and become habitable remains uncertain. The stellar flux required to melt these planets will vary as a function of host star spectral type due to the wavelength dependence of the ice albedo \citep{shields2013,shields2014}. In some cases the required flux to thaw a planet may be high enough to lead directly to a runaway greenhouse state \citep{Yang2017}, while planets able to outgas thicker CO$_2$ atmospheres may have an easier time deglaciating \citep{Wolf2017}.

Given that BHZ planets are located in the habitable zone, but may not actually be habitable, it is important to determine whether planets that we discover will belong to this class of planets. In order to determine whether a given planet found in the habitable zone is in the BHZ, one requires knowledge of the evolution of its host star and corresponding habitable zone. \citet{Tuchow2021} demonstrate that one can easily determine if a planet is in the BHZ or sustained CHZ given models of the habitable zone and stellar structure and evolution, provided one has knowledge of the stellar fundamental properties of mass, age, and metallicity. 
With the ability to calculate the boundaries of the BHZ given a host star's properties, 
one would like to understand how common these ambiguously habitable BHZ planets are around different types of stars. In planning for future missions, it is important to assess how estimates of science yields will be affected by our current uncertainty about the habitability of planets. In this study, we will investigate how abundant these planets are around different types of stars, and we will determine how different formulations of the CHZ lead to different populations of stars being preferred in terms of the frequency of CHZ planets. We would like to know how different definitions of continuous habitability may bias the target selection for future missions to directly image and characterize the atmospheric compositions of planets. One would like to determine how the ambiguity in the habitability of BHZ planets will affect the ability of future direct imaging missions to find biosignatures.

\section{The Occurrence Rates of Belatedly Habitable Planets}
\label{BHZ_occurrence}

In this study, we will take an agnostic approach to the habitability of BHZ planets, and treat them as a known unknown when planning for future missions. To estimate the science outputs of future space-based direct imaging missions, we would like to estimate the occurrence rates of planets in the sustained CHZ and by extension in the BHZ. CHZ planets may pose the best chances to host biosignatures, while BHZ planets pose an intriguing test for our theories of planetary habitability, and future missions to image habitable zone planets may be able to empirically test their habitability. 

For estimation of the occurrence rates of BHZ planets, we will first determine the fraction of habitable zone planets that lie in the BHZ as a function of host star fundamental properties.
We start off by generating a grid of stellar models in mass and age. We use MIST isochrones and stellar tracks \citep{Dotter2016,Choi2016}, as well as the \texttt{isochrones} python package \citep{Morton2015} to calculate for each grid point how a star's luminosity and effective temperature change in time. This informs the evolution of the habitable zone, in this case calculated using the \citet{Kopparapu2013} formulation.
Knowledge of a star's evolutionary track allows us to determine which regions of the habitable zone belong to the belatedly habitable zone. Beyond just knowing the spatial extent of the belatedly habitable zone, we would like to determine the fraction of planets in the habitable zone that fall in the BHZ. This requires us to have an understanding of the distribution function for rocky planets in the habitable zone, $\Gamma(a)$. 
The planetary distribution function is not very well constrained in regions of parameter space around the habitable zone. Due to the small sample size of known planets at these separations, past attempts to characterize the distribution function relied on extrapolations to longer periods \citep{Petigura2013} or inferences based on limited data points \citep{Burke2015}. Furthermore, $\Gamma$ appears to depend on additional variables such as host star mass \citep{Neil2018}, though sample sizes are not typically large enough to measure this dependence except in the case of very large mass bins, comparing for example the distribution of planets around M stars versus that around FGK stars \citep{Hsu2019,Hsu2020}.
%more on planetary distribution function here
Given the uncertainties in determining the planetary distribution function, for the purposes of this study, we will consider the simple approximation that planets are uniformly distributed in log semimajor axis. This planetary distribution implies that planets around other stars would be similar to the distribution of planets in the solar system (i.e. Bode's Law), and this distribution is actually fairly close to our best estimates for a split powerlaw fit in this region of parameter space \citep{Dulz2020}. The fraction of habitable zone planets, $f_{\mathrm{BHZ}}$, in the BHZ is given by 
\begin{equation}
    f_\mathrm{BHZ} =  \frac{\int_\mathrm{BHZ} \Gamma(a) \,da}{\int_\mathrm{HZ} \Gamma(a) \,da} = 1 -\frac{\int_\mathrm{CHZ} \Gamma(a) \,da}{\int_\mathrm{HZ} \Gamma(a) \,da}
\end{equation}
where $a$ is the semimajor axis and integrals are over the inner and outer bounds of the HZ, CHZ, and BHZ respectively. The normalization of the planetary distribution function $\Gamma(a)$ is unimportant as it cancels out in the equations. As shown in the equation above, from knowledge of $f_{\mathrm{BHZ}}$ it is straightforward to calculate the fraction of HZ planets in the sustained CHZ as $1 - f_\mathrm{BHZ}$. 
The code used for our CHZ calculations can be found at the publicly available \texttt{CHZ\_calculator} repository\footnote{\url{https://github.com/nwtuchow/CHZ_calculator.git}}.
%mention choice of Gamma is a rough estimate

Determining the position of the BHZ requires one to select a time at which habitability is considered to start, $t_0$, from which sustained CHZ planets have remained in the habitable zone and after which BHZ planets have entered the habitable zone. Determining the time for the onset of habitability and the emergence of surface liquid water is a complex and multifaceted area of research. Rather than trying to determine the exact time at which planets become habitable, in this study we will consider the arrival of volatiles as a necessary condition, and therefore a good proxy for the onset of habitability. 

Since the time required for the arrival of volatiles is not known and will vary greatly for most systems, in this study we consider two timescales for the delivery of volatiles, $t_0$, that bookend realistic values for a wide range of stars. 
First we consider the case where habitability begins after the delivery of volatiles at $t_0 = 10$ Myr --- a very early time for the delivery of volatiles, corresponding with the lifetime of the protoplanetary disk \citep{Ribas2014}. This timescale for the onset of habitability would imply that planets obtain the majority of their volatiles from their formation in the protoplanetary disk, and may risk obstacles to eventually becoming habitable shortly after the dissipation of the disk. The timescale for the formation of planetary embryos can vary widely between models. For scenarios where it takes longer for planetary embryos to form, much of the accretion of volatiles would occur after 10 Myr, after the dissipation of the protoplanetary disk \citep{raymond2006}. Our opposite bookend for the delivery of volatiles is that planets do not acquire the majority of their water until after $t_0 = 100$ Myr, corresponding to a scenario where most of a planet's volatiles originate via impacts and Late Veneer accretion \citep{brasser2016}. It also corresponds to a high estimate for the timescale for the formation of a solid surface for rocky planets in the habitable zone \citep{Hamano2013}.

\begin{figure}
    \plottwo{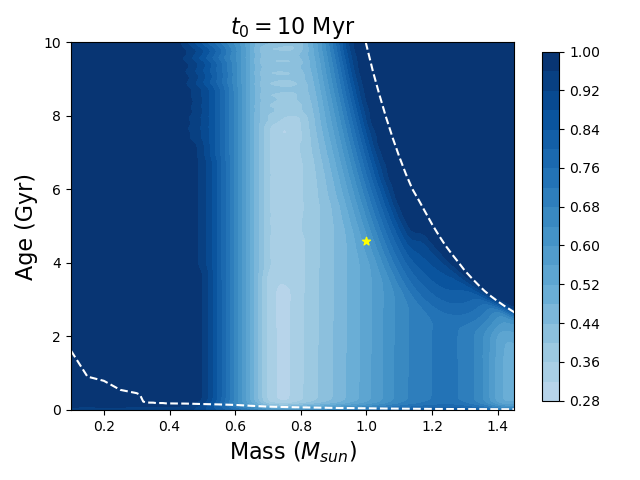}{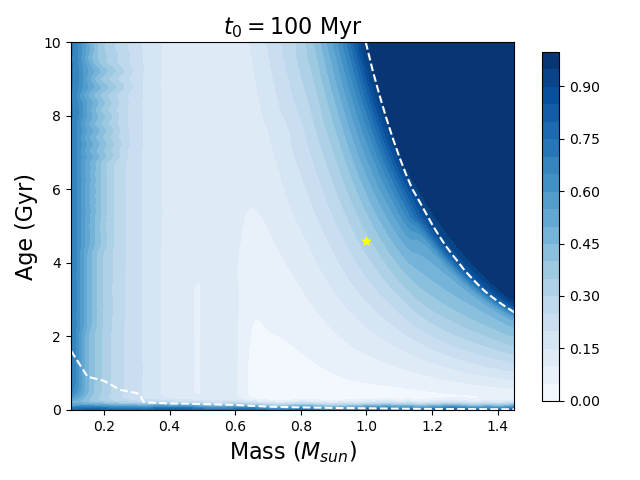}
    \caption{Fraction of habitable zone planets in the BHZ as a function of host star mass and age. Here we show a cross section of stellar mass, age, and metallicity space taken at solar metallicity of [Fe/H] =0.0. Left panel shows the case for an onset of habitability after $t_0 = 10$ Myr, while the right panel shows a $t_0 = 100$ Myr onset of habitability. The yellow star icon shows the position of the sun, while the dashed white lines represent the zero age main sequence and terminal age main sequence.}
    \label{BHZ_planet_fraction}
\end{figure}

In Figure \ref{BHZ_planet_fraction} we plot the fraction of habitable zone planets in the BHZ as a function of the mass and age of the host star. In this plot, metallicity has been held at the solar value of [Fe/H] = 0.0, but we also compute the BHZ planet fraction for all the grid points in metallicity space given by the MIST isochrones.  The masses and ages represented in this figure have been selected to correspond with the range of masses and ages of host stars which one would expect to observe in a search for biosignatures. One is unlikely to observe stars with ages greater that 10 Gyr, the age of the galactic disk \citep{Carraro2000}, and stars beyond spectral type A would likely evolve too quickly to allow for the emergence of detectable life.

The left panel shows the case for an early onset of habitability after $t_0 = 10$ Myr. One can see in this plot that sun-like stars have on average close to 60\% of their planets in the BHZ. Stars around 0.7 $M_\odot$ have the lowest fraction of BHZ planets (i.e. the highest fraction of sustained CHZ planets), while lower mass stars appear to have almost the entirety of their habitable zones occupied by BHZ planets. This makes sense as M dwarfs spend up to hundreds of millions of years contracting and dimming on the premain sequence and early main sequence, so having an early onset of habitability means that most of this dimming happens after the onset of habitability. One can also see in this plot that older stars have a higher fraction of BHZ planets. For more massive stars with shorter lifetimes, one can see that prior to the terminal age main sequence (represented by the upper white dashed line), they reach a point where all of the planets in their habitable zone are BHZ planets.

The right panel of Figure \ref{BHZ_planet_fraction} shows the more optimistic case where habitability is considered to start much later, after $t_0 = 100$ Myr. In this case, most of the early dimming of lower mass stars has already occurred, so a smaller fraction of their planets are in the BHZ. We can still observe the behavior where, as stars of a given mass evolve, the fraction of BHZ planets increases. In this scenario the habitable zones of sun-like stars are somewhat evenly split with around 45\% of planets in the BHZ.

For future missions to search for biosignatures, we would like to estimate the fraction of habitable zone planets they will observe that will be BHZ planets. To gain a rough estimate of the fraction of BHZ planets we generate a random sample of 10000 target stars and determine the mean fraction of BHZ planets. We draw these stars' masses and metallicities (in terms of [Fe/H]) from the known distribution of stellar properties for nearby bright stars that will be the targets of future exoplanet direct imaging missions using the ``Mission Stars" catalog\footnote{\url{https://exoplanetarchive.ipac.caltech.edu/docs/MissionStellar.html}} hosted by the NASA Exoplanet Archive \citep{Turnbull2015}. As most of these target stars lack precise age estimates, we draw ages from a uniform distribution between 200 Myr and 10 Gyr (avoiding very young stars). For each star in the sample, after generating a mass and an age, we calculate the fraction of BHZ planets, interpolating the values calculated for the stellar model grid seen in Figure \ref{BHZ_planet_fraction}. We find that the average fraction of habitable zone planets that are in the BHZ is $\overline{f}_\mathrm{BHZ} = 0.74$ for $t_0 = 10$ Myr and $\overline{f}_\mathrm{BHZ} = 0.29$ for $t_0 = 100$ Myr. Note that the values for the fraction of BHZ planets which we calculate here should be treated as rough estimates owing to the poorly constrained planetary distribution function $\Gamma$ in these regions of parameter space as well as the uncertainties in the onset of habitability and volatile delivery. Nonetheless, for reasonable forms of $\Gamma$ and values for $t_0$ one would expect $\overline{f}_\mathrm{BHZ}$ to fall between the two values above.

It is clear that the time for the delivery of volatiles and the onset of habitability plays a major role in determining the fraction of BHZ planets that future missions would expect to observe. Regardless of the correct value for $t_0$, between 29--75\% of HZ planets will reside in the BHZ. This would comprise a large portion of potentially habitable planets found by future missions, and assessing whether these planets can be habitable is of critical importance when assessing the scientific yield of such missions, and whether they could reasonably expect to detect biosignatures. 
%One should note that our calculations of the fraction of BHZ planets are coarse estimates based on our limited knowledge of the planetary distribution in the habitable zone, the emergence of habitable conditions on exoplanets, and the delivery of volatiles. Nonetheless, our estimated range of values for the 
%rephrase

\section{Differences in Continuous Habitability}
\label{CHZ_differences}

\begin{figure}
    \plottwo{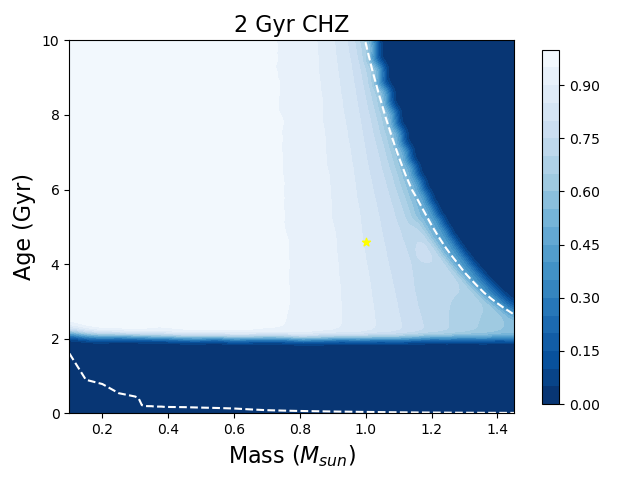}{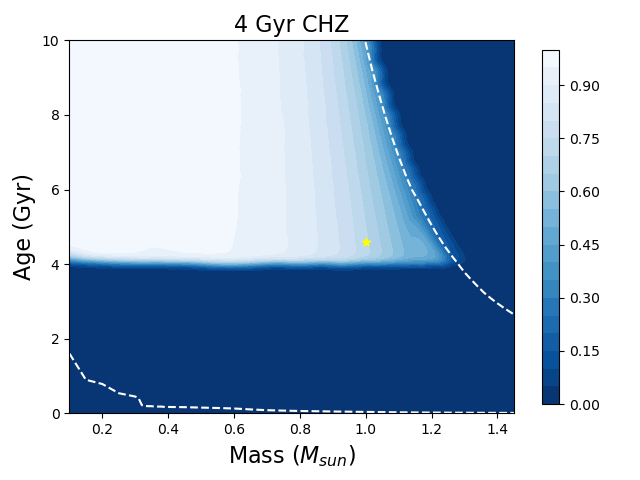}
    \caption{Fraction of habitable zone planets in the fixed duration CHZ. Here we show a cross section of stellar mass, age, and metallicity space taken at solar metallicity of [Fe/H] =0.0. The left panel shows the case 2 Gyr as the duration required for continuous habitability, while the right panel requires 4 Gyr for continuous habitability.}
    \label{alt_CHZ_fp}
\end{figure}

Another way to interpret the results of Figure \ref{BHZ_planet_fraction} is to see the fraction of BHZ planets as one minus the fraction of planets in the habitable zone that are in the sustained CHZ. Regions of these plots with the lowest fractions of planets in the BHZ correspond to stars with the highest fraction of planets in the sustained CHZ.  However, one should note that differences in models and assumptions about planetary habitability result in different formulations of the CHZ.
We are interested in investigating how different definitions of the CHZ lead to different populations of stars being preferred in a search for life. In Figure \ref{alt_CHZ_fp}, we plot the fraction of planets in the alternative, fixed duration CHZ for the same stellar model grid seen earlier. Here we consider the cases where planets are considered continuously habitable after remaining in the habitable zone for 2 Gyr and 4 Gyr in the left and right panels respectively. In this plot, the color scale has been chosen to be the opposite of that in Figure \ref{BHZ_planet_fraction}, so that the fraction of planets in the CHZ is more readily comparable between the two figures. 

%edit this cutoff changed in most recent figure
One can see that the most noticeable feature in this plot of the fixed duration CHZ is the sharp cutoff between stars with no fixed duration CHZ and stars where a large fraction of their habitable zone is occupied by the fixed duration CHZ. This cutoff typically occurs around the fixed age required for continuous habitability, with small variations due to the premain sequence evolution of stars. This sharp cutoff is non-physical in nature and planets that have spent a little less time in the habitable zone than the fixed duration for continuous habitability are unlikely to be significantly worse candidates to host life. 
One benefit of this habitable zone formulation is that it gives preference to planets that have been in the habitable zone for a longer duration. The sustained CHZ only considers planets that have remained in the habitable zone since their formation, not the time that they have stayed in the habitable zone. This means that almost all planets around young stars are in the sustained CHZ by this definition, but this does not necessarily mean they are good locations in the search for life. 

Another major difference between the sustained and fixed duration definitions of continuous habitability is in their treatment of lower mass stars. In the early onset of habitability case ($t_0=10$ Myr) for the sustained CHZ, all planets in the habitable zones of the lowest mass stars are BHZ planets, and there is no CHZ. Contrast this with the fixed duration CHZ, where past a certain stellar age, low mass stars have almost all of their habitable zones occupied by CHZ planets. One can see that, depending on the definition of the CHZ that is used, low mass stars could either be the best or the worst candidates in a search for biosignatures. The situation is not as severe in the $t_0=100$ Myr scenario, but the fraction of planets in the CHZ still varies significantly between the two cases for low mass stars.

Judging if a planet is a good candidate to host biosignatures based on whether it resides in the CHZ, appears to have major pitfalls. For one, the different definitions of the CHZ vary substantially between stars of different masses and ages. A given planet can fall in the CHZ by one definition, and outside of it by another. Both definitions of the CHZ have their shortcomings. The sustained CHZ only determines whether a planet has been consistently in the habitable zone up until the current day, and doesn't incorporate knowledge about how long planets spend in the habitable zone. The fixed duration CHZ on the other hand incorporates the time spent in the habitable zone, but it includes a non-physical sharp age cutoff, and doesn't take into account where planets are belatedly habitable. 

\subsection{Beyond current definitions of the CHZ}
\label{metric_section}
The different definitions of the continuously habitable zone appear to conflict with each other and disagree about which stellar populations are the best host stars for planets with long-term habitability. These contrasting formulations of the CHZ prioritize either the continuity of habitable conditions or the duration spent in the habitable zone, but realistically both of these factors are important considerations when considering the long-term habitability of exoplanets. Both definitions of the CHZ appear to be incomplete, and one should take into account both the duration which a planet has been habitable and whether a planet has been consistently habitable since the onset of habitability.  
One could potentially develop a new definition of the CHZ which incorporates both of these factors, but it is potentially more useful to take a step back and rethink the usefulness of the concept of a continuously habitable zone. Defining a CHZ is akin to assigning a Boolean value as to whether a planet meets a given criterion relating to long-term habitability. Planets within the CHZ would meet this criterion and planets outside it would not. However, one should note that some planets in the CHZ may be better candidates to host life and biosignatures than others. Therefore it would be useful to define a continuous quantity to assess the quality of a target to host long-term habitability and biosignatures.

This was the motivation for \citet{Tuchow2020} to develop a framework for biosignature yield metrics. Given knowledge of a host star's fundamental properties and evolutionary track, one can determine whether planets in its habitable zone remain habitable for an extended duration. \citet{Tuchow2020} define their metrics as 
\begin{equation}
    B = \int H(a,t) \Gamma(a)\,da
    \label{metric_eq}
\end{equation}
where $\Gamma$ is the planetary distribution function as seen earlier and $H$ is a quantity related to the probability of biosignature emergence. Both of these quantities are based on our prior knowledge and assumptions about the distribution of habitable zone planets and the emergence of life. They allow one to compute a quantity $B$ which can serve as a prior for which stars would be the best hosts for long term habitability. $B$ is calculated as a property of a host star, and realistically it can be used in much the same way as the fraction of habitable zone planets in the CHZ, described in earlier sections. Note that this equation is a marginalized form of the double integral for $B$ in \citet{Tuchow2020}, integrating over planetary radius so the only free dimension in the integrand is semi-major axis, $a$.
%While these metrics $B$ were designed with the emergence of biosignatures in mind, realistically they can be used to assess the same considerations for long-term habitability as are included in the

\begin{figure}
    \centering
    \plotone{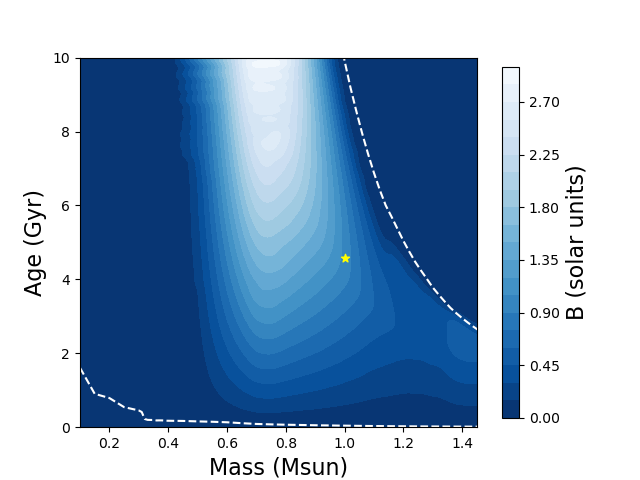}
    \caption{Example $B$ metric including elements of both the sustained CHZ and fixed duration CHZ. Here the time for the delivery of volatiles is set to 10 Myr. $B$ values are normalized to the value calculated for the current day sun.}
    \label{example_metric}
\end{figure}

The framework of \citet{Tuchow2020} can be readily applied to construct a metric that takes into account elements of both definitions of the CHZ. Using Equation \ref{metric_eq}, we select $H$ and $\Gamma$ to best incorporate the elements of long-term and sustained habitability. Here we choose to use the same power law form of the planetary distribution function, $\Gamma(a)$, as we use in Sections \ref{BHZ_occurrence} and \ref{CHZ_differences}. The quantity $H(a,t)$ contains all of the information relating to the long-term habitability of exoplanets. It can be interpreted as a weight applied to the planetary distribution function, assessing the quality of a planet as a target given its age, and distance from its star. For this metric, we use a form of $H$ that incorporates elements of both CHZ definitions by defining
\begin{equation}
    H = 1 - e^{-b\tau(a,t)}
\end{equation}
where $\tau(a,t)$ is the duration that a planet at separation $a$ and age $t$ has spent in the sustained CHZ (defined with a time for volatile delivery of 10 Myr), and $b$ determines how strongly the duration spent in the habitable zone is prioritized. If we choose to interpret $H$ as the probability that a given planet hosts biosignatures, then $b$ would be the rate at which biosignatures emerge per unit time. Here we use a value of $b$ = 1/30 Gyr$^{-1}$, but we note that for small $b\tau$, which is the case for most stars, the relative values of $B$ are insensitive to $b$ (see Section 2.3 in \citet{Tuchow2020}). The metric described above gives preference both to planets that have been in the habitable zone since the delivery of volatiles, and planets that have stayed in the habitable zone for an extended duration. It therefore combines both definitions of the CHZ into a quantity $B$ representing our understanding of the importance of long-term and sustained habitability.

For the grid of stellar models used in earlier sections, we calculate this metric, $B$, for each grid point and interpolate it in terms of stellar mass, age, and metallicity. In Figure \ref{example_metric} we plot the values of our example metric as a function of stellar mass and age for the case of solar metallicity [Fe/H]=0.0. We normalize the values of this metric to the value calculated for the current day sun, and observe how the relative values of this quantity compare between stars in different regions of parameter space. One can see that our example $B$ metric behaves somewhat like a combination of the fractions of CHZ planets shown in Figures \ref{BHZ_planet_fraction} and \ref{alt_CHZ_fp}. It favors both stars with older ages, and stars whose habitable zones have remained stable since the delivery of volatiles. This means that very low mass stars, that dim significantly after the delivery of volatiles, as well as younger stars and more evolved stars have lower values for this metric. The highest values for this metric correspond to older stars of stellar masses between 0.5 and 0.9 $M_\odot$, which would likely be K stars or early M stars. This region of parameter space similarly contains a large fraction of habitable zone planets in the CHZ for both CHZ formulations, so the metric we have constructed serves to demonstrate the overlap in the stellar populations preferred by both definitions. With our understanding of importance of the time dimension of habitability, this metric suggests that older K stars would be the best targets in terms of the stability of their habitable zones and the duration which planets spend in the habitable zone.   

\section{Discussion and Conclusions}

To determine whether a planet is a likely candidate to host biosignatures, it is important to consider whether it has remained consistently habitable throughout its lifetime. This concept of sustained habitability is included in some definitions of the continuously habitable zone, and in this study we use the term belatedly habitable planets to refer to planets that entered the habitable zone from outside due to the evolution of their host stars. We have seen in Section \ref{BHZ_occurrence} that between 29 - 74\% of planets in the habitable zone will belong to this class of BHZ planets. There is substantial uncertainty in estimating the fraction of planets in the BHZ due to our uncertainty in determining the time for the onset of habitability and the delivery of volatiles, but considering contrasting edge cases we can bound the fraction of BHZ planets within this range.  Even the lower bound of this estimate would imply that a huge portion of planets in the habitable zone are BHZ planets. We can expect that many exoplanets that will be found in the habitable zones of other stars will belong to this class of BHZ planets,  especially in the case of lower mass stars, where most planets will likely be BHZ planets.

While they reside in the current day habitable zone, it remains unclear as to whether BHZ planets can truly be considered habitable. These planets face potentially greater obstacles to habitability than planets in the sustained CHZ, and future planetary modelling efforts are required to determine whether these planets can retain their liquid water and host temperate surface conditions, and if so under which conditions.  As a large fraction of planets in the habitable zone will be ambiguously habitable BHZ planets, this raises the question as to whether one can infer if a planet is habitable based solely on if it lies in the habitable zone.

Beyond sustained habitability and whether a planet is belatedly habitable, it is important to consider the duration that planets have maintained habitable conditions. In the case of Earth, it took billions of years of evolution for complex life to emerge, and for oxygen to reach atmospheric concentrations sufficient for it to be detectable to a remote observer. If life on other planets evolves over similar timescales, then, in the search for life, it makes sense to prioritize planets that have been habitable for billions of years. Several definitions of continuous habitability, what we call the fixed duration CHZ formulation, define the CHZ as a region occupied by planets that have spent more than a given duration in the habitable zone. In Section \ref{CHZ_differences}, we calculate the fraction of habitable zone planets in the fixed duration CHZ as a function of stellar mass and age for different required durations for continuous habitability. We find that these formulations of the CHZ give preference to older stars in terms of the fraction of CHZ planets, but also introduce a sharp cutoff in stellar age based on these definitions of the CHZ. For instance, using the fraction of planets in the 2 Gyr fixed age CHZ, a star with an age of 2.05 Gyr could be a great target in a search for life, whereas a stellar age of 1.95 Gyr would result in a star having no potential to host planets in the CHZ. This may pose a major problem, as stellar ages are often notoriously difficult to constrain \citep{Soderblom2010}. Age determination is particularly difficult for lower mass stars on the main sequence, as they evolve very slowly, such that stars with ages billion of years apart appear to be at similar evolutionary stages. For typical field stars, such as those in the \textit{GAIA-Kepler} Stellar Properties Catalog, one can expect high median age uncertainties on the order of 56\% \citep{Berger2020a}. Even for the best target stars with independent age constraints from gyrochronology, one can typically only achieve a median age uncertainty of 14\% for cool dwarf stars \citep{Claytor2020}. The large uncertainty in stellar ages makes it difficult to determine whether a planet falls in the fixed duration CHZ. 
Many stars will straddle the sharp age cutoff introduced by the fixed duration formulation of the CHZ, making their fraction of CHZ planets highly uncertain. Rather than using a CHZ formulation with a sharp age cutoff, it may instead be preferable to use a metric such as that introduced in Section \ref{metric_section}, that gives higher values for longer durations spent in the habitable zone. Such a metric could be useful in the comparison of systems in terms of their likeliness of hosting life and biosignatures, and could propagate our uncertainties in stellar ages and other host star properties (see for example the analysis of \citet{Tuchow2022}).

Perhaps the most concerning difference in the stellar populations preferred in terms of the fraction of CHZ planets in the habitable zone is for the case of low mass stars. Depending on the definition of the CHZ that one uses, older low mass stars could either host the highest fraction of CHZ planets or host none at all. This is due to the fact that planets around M dwarfs form during the early dimming of the star on the premain sequence. This means that many planets currently in the habitable zones of these stars are inner BHZ planets, formerly too hot to support habitable conditions, which may have been susceptible to a substantial loss of volatiles early in their histories. For older stars, such planets would remain in the habitable zone for an extended duration after entering it, but whether they would actually be able to host surface liquid water given their extreme initial conditions remains unclear. The different definitions of the CHZ struggle to categorize these planets as either being poor or excellent locations for continuous habitability based on whether duration in the habitable zone or sustaining habitability from formation are prioritized.

The conflicts we have observed between different definitions of the CHZ raise the question as to whether the CHZ is still useful as a concept, and if so how can it best be applied. Despite the disagreement between the different formulations of the CHZ, both highlight important areas when studying the long-term habitability of exoplanets. When searching for biosignatures, both whether a planet has remained consistently habitable -- emphasized by the sustained CHZ -- and how long a planet has maintained habitable conditions -- emphasized by the fixed duration CHZ -- are important considerations. Both definitions assess elements of a planets habitable history, but neither is able to give the full picture.  
Because of this, we believe that, in their current form, it would be unwise to use either of these definitions of the CHZ as a means to prioritize targets for future biosignature surveys, as both are too dependent on their assumptions about currently poorly understood areas of planetary habitability. 

The concept of the CHZ provides a useful heuristic to assess the potential for planets to host life and biosignatures, but it is incomplete and could be improved.
Perhaps a more useful expansion upon the concept of the CHZ could incorporate elements of both competing CHZ formulations, assessing both whether planets have been consistently habitable and for what duration.
The biosignature yield framework of \citet{Tuchow2020} aimed to integrate both these aspects of long-term habitability and a model for biosignature emergence to assess which stars are most likely to host planets with biosignatures. In Section \ref{metric_section}, we constructed an example long-term habitability metric that combines the considerations of both CHZ definitions. For target star prioritization in a search for biosignatures, one could use this metric or develop a Bayesian framework along these lines to assess how likely a given planet is host habitable conditions or signs of life. Alternatively, future missions could be more agnostic as to which targets to observe, not taking an overly restrictive approach, and searching beyond the currently calculated boundaries of the habitable zone or CHZ. 

The CHZ represents our current understanding of where we would expect to find biosignatures, but rather than biasing our surveys to only observe CHZ planets, we should instead treat it as a hypothesis to be tested empirically by future missions.
With future instruments, astronomers aim to measure the spectra of planets in the habitable zone and infer their atmospheric compositions and whether they host habitable conditions or signs of life. 
With a large enough sample planetary spectra, we hope to be able to take a statistical comparative planetology approach and test hypotheses about the exoplanet population \citep{Bean2017}. In particular, we would like to study remote indicators of planetary habitability, such as the dependence of atmospheric CO$_2$ concentrations on stellar incident flux \citep{Checlair2019,Lehmer2020}, and determine empirically where the boundaries of the habitable zone lie.
We would like to determine which types of planets habitability can appear on, and observe whether BHZ planets would be able to host habitable conditions. In the near future, the James Webb Space Telescope will likely be able to obtain transmission spectra for planets in the habitable zones of M stars. One of the best candidates for JWST observation is the Trappist-1 system which possesses multiple rocky planets in the habitable zone \citep{lustig_yaeger2019}. Due to the fact that M stars dim early in their evolution, the planets currently in the habitable zone of Trappist-1 are almost certainly inner BHZ planets. This means that they were previously located interior to the habitable zone, and may have experienced large amounts of water loss prior to entering the habitable zone \citep{Luger2015}. These planets will be an ideal observational test bed for our theories about inner BHZ planets, and will allow us to empirically test whether these planets are able to retain their volatiles and host habitable conditions.

In the more distant future, astronomers aim to characterize the atmospheric compositions of rocky planets in the habitable zones of sun-like stars using direct imaging methods. One of the top priorities of the Astro2020 Decadal Survey is the direct imaging of Earth sized planets in the habitable zone \citep{DecadalSurvey}, and mission concepts such as HabEx and LUVOIR demonstrate that this precision would be achievable \citep{HabEX_Final_report,LUVOIR_final_report}.
Due to the nature of exoplanet direct imaging observations, planets at larger separations from their host stars are easier to observe because the host star's light can be more readily blocked out without obscuring the planet's. This means that, in term of planets in the habitable zone, BHZ planets would be easier to observe using direct imaging methods. As the habitability of BHZ planets is ambiguous, these future missions will greatly benefit by not assuming that these planets are habitable. 

While definitions of continuous habitability can conflict with each other, the long-term habitability of exoplanets is of vital importance when planning for future biosignature surveys. We would like to determine whether biosignatures are restricted to planets in the CHZ (the boundaries of which vary by definition) or if BHZ planets could also be habitable for an extended duration and have a chance of hosting life. Given that a large portion, maybe even the majority of planets in the habitable zone will be BHZ planets, the question of their habitability has wide reaching implications for future mission design, and will influence which targets these missions observe and what search strategies are employed. 
%In order to determine whether planets surveyed by future missions belong to the classes of BHZ planets or CHZ planets would require one to obtain precise masses, ages and metallicities of host stars in order to determine their evolutionary tracks.

%Belatedly habitable planets form a unique class of planets characterized by their host star's evolution. With future missions and advancements in technology, we are just now attaining the precision to characterize assess the habitable histories of exoplanets and to test whether such planets are habitable.

\section*{Acknowledgements}

NWT is supported by an appointment with the NASA Postdoctoral Program at the NASA Goddard Space Flight Center, administered by Oak Ridge Associated Universities under contract with NASA.
The Center for Exoplanets and Habitable Worlds and the Penn State Extraterrestrial Intelligence Center are supported by the Pennsylvania State University and the Eberly College of Science.

This research has made use of NASA's Astrophysics Data System Bibliographic Services as well as the NASA Exoplanet Archive, which is operated by the California Institute of Technology, under contract with the National Aeronautics and Space Administration under the Exoplanet Exploration Program.
This work used the \texttt{Isochrones} (\url{https://github.com/timothydmorton/isochrones}) python package for its calculations.
The code used for the calculations in this study can be found at \url{https://github.com/nwtuchow/CHZ_calculator.git}.

\facility{Exoplanet Archive}

\bibliographystyle{aasjournal}
\bibliography{sources}

\end{document}